\documentclass[conference,a4paper]{IEEEtran}
\IEEEoverridecommandlockouts
\usepackage{cite}
\usepackage{amsmath,amssymb,amsfonts}
\usepackage{algorithmic}
\usepackage{graphicx}
\usepackage{textcomp}
\usepackage{xcolor}
\usepackage{subfig}
\usepackage{url}
\usepackage[hidelinks]{hyperref}
\hypersetup{breaklinks=true}
\def\BibTeX{{\rm B\kern-.05em{\sc i\kern-.025em b}\kern-.08em
    T\kern-.1667em\lower.7ex\hbox{E}\kern-.125emX}}
\begin{document}

\title{Design and Implementation of a Real-Time Multi-Site Immersive Learning System Using Photon Fusion}

\author{\IEEEauthorblockN{Iwai Wataru, Duc V. Nguyen}
\IEEEauthorblockA{\textit{Department of Communication and Information Engineering} \\
\textit{Tohoku Institute of Technology}\\
Sendai, Japan}
}

\maketitle

\begin{abstract}
In this paper, we develop a Virtual Reality-based immersive learning environment that allows teachers to conduct a lesson in a virtual space using Photon Fusion. The proposed system allows teachers and students to be present in the same virtual space regardless of their actual physical locations. The teachers can verbally communicate with students in real-time, interacting with 3D learning materials. By adopting Photon Fusion, the system achieves stable real-time communication and synchronization among multiple players. Evaluation results demonstrate that the proposed system provides stable communication performance, good usability, and minimal VR sickness, confirming its effectiveness as an immersive learning environment. 
\end{abstract}


\section{Introduction}

In recent years, with the widespread popularity of VR devices such as the Meta Quest 3, educational methods using VR have received significant attention. Because VR offers a high degree of immersion, it enables students to experience learning activities in a virtual space that are difficult to simulate in traditional classroom settings, such as gaining a three-dimensional understanding of human structures or conducting chemistry experiments without being constrained by the physical environment.

In~\cite{IwaiGCCE2024}, the authors have previously proposed an immersive learning environment that allows teachers and students to participate in the same virtual space and conduct classes in real time. In addition to two-way voice communication, this system enables the sharing and manipulation of 3D models, thereby providing an immersive and interactive learning environment. 
However, the system proposed in~\cite{IwaiGCCE2024} has several issues. The first is the stability of the information sharing system. In~\cite{IwaiGCCE2024}, a server was run on a PC, and WebSocket and WebRTC were used to send and receive 3D models and audio. However, since the transmission of information was conducted via peer-to-peer, the performance of the system depended on the processing power of the Head-Mounted Display (HMD). Consequently, when multiple users participated simultaneously, the processing load increased, making the frame rate (FPS) unstable and making smooth operation difficult. Therefore, only up to five people could participate. In~\cite{IwaiGCCE2024}, issues with the stability of the information-sharing system, combined with complex operation methods such as hand tracking, resulted in poor usability.

To address these challenges, this study aims to develop an immersive VR-based learning environment that features a highly stable information-sharing system supporting twenty participants and achieves high usability through improved operability. 



\begin{figure}[htbp]
  \begin{center}
    \includegraphics[width=\columnwidth]{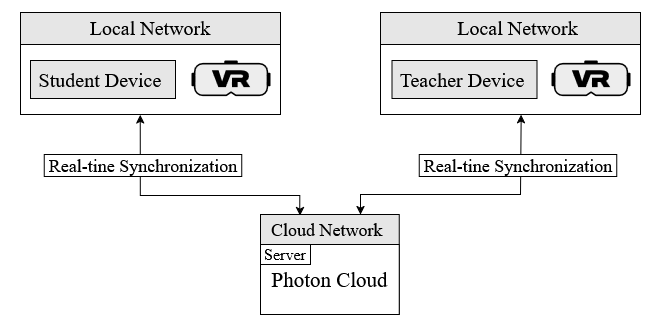}
  \end{center}
  \caption {System configuration diagram of the proposed system}
  \label{SystemsConfigurationDiagram}
\end{figure}

\section{Proposed System}

Fig.~\ref{SystemsConfigurationDiagram} shows the configuration diagram of the proposed system. Students and teachers connect to the Internet via Wi-Fi using a VR headset and communicate through Photon Cloud, which acts as a central server to synchronize object states and audio among all clients in real time ~\cite{photon}. 
The proposed system’s real-time voice communication is implemented by integrating Photon Fusion (the network engine) with Photon Voice 2 (the audio processing library). This allows us to establish a mechanism where voice communication is initiated in conjunction with the client’s entry into the VR classroom.



\begin{figure}[htbp]
  \begin{center}
    \includegraphics[width=0.9\columnwidth]{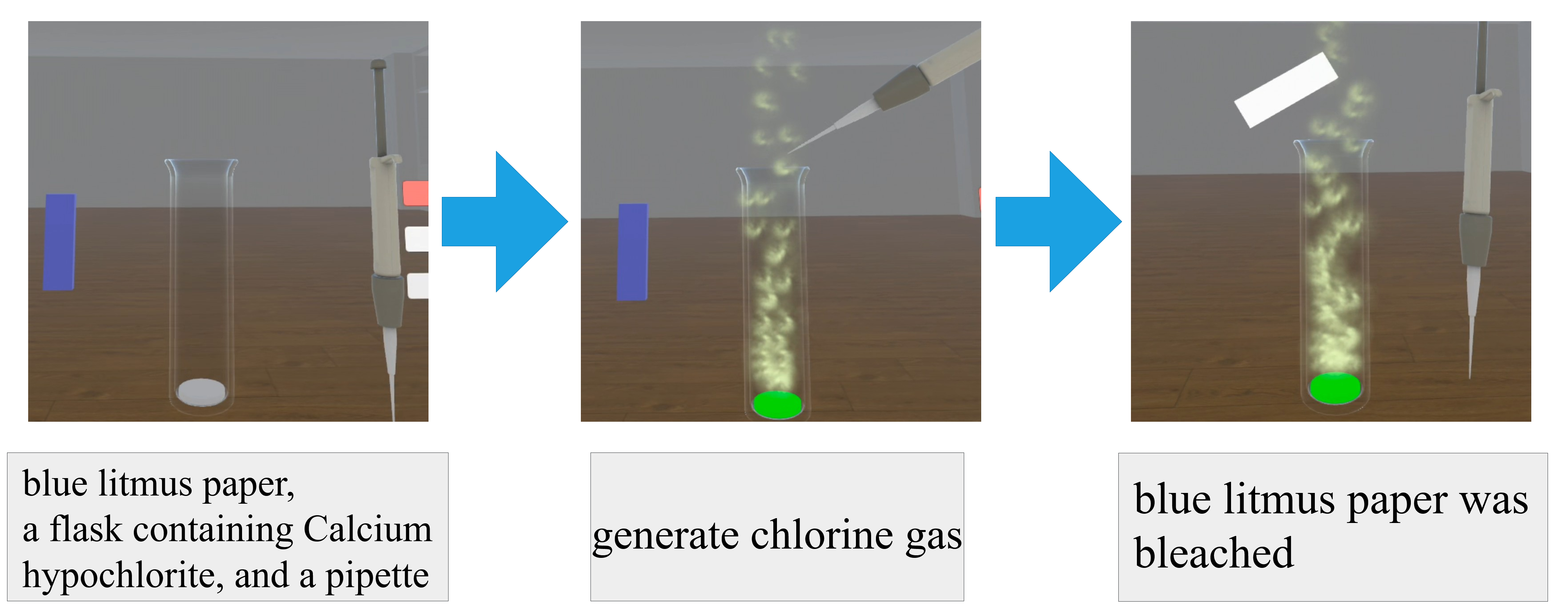}
  \end{center}
  \caption{Learning materials of the proposed system}
  \label{LearningMaterials}
\end{figure}

Fig.~\ref{LearningMaterials} shows the learning materials used in the experiment and their operation. The tools are shown in the image on the left in Fig.~\ref{LearningMaterials}. From left to right, they are blue litmus paper, a flask containing Calcium hypochlorite, and a pipette containing hydrochloric acid. The center of Fig.~\ref{LearningMaterials} shows the process of generating chlorine gas by manipulating the pipette. The chlorine gas is rendered as yellow-green smoke. In the experiment, the blue litmus paper bleaches when it comes into contact with chlorine gas. The image on the right in Fig.~\ref{LearningMaterials} shows this phenomenon.

\begin{figure}[htbp]
  \begin{center}
    \includegraphics[width=0.9\columnwidth]{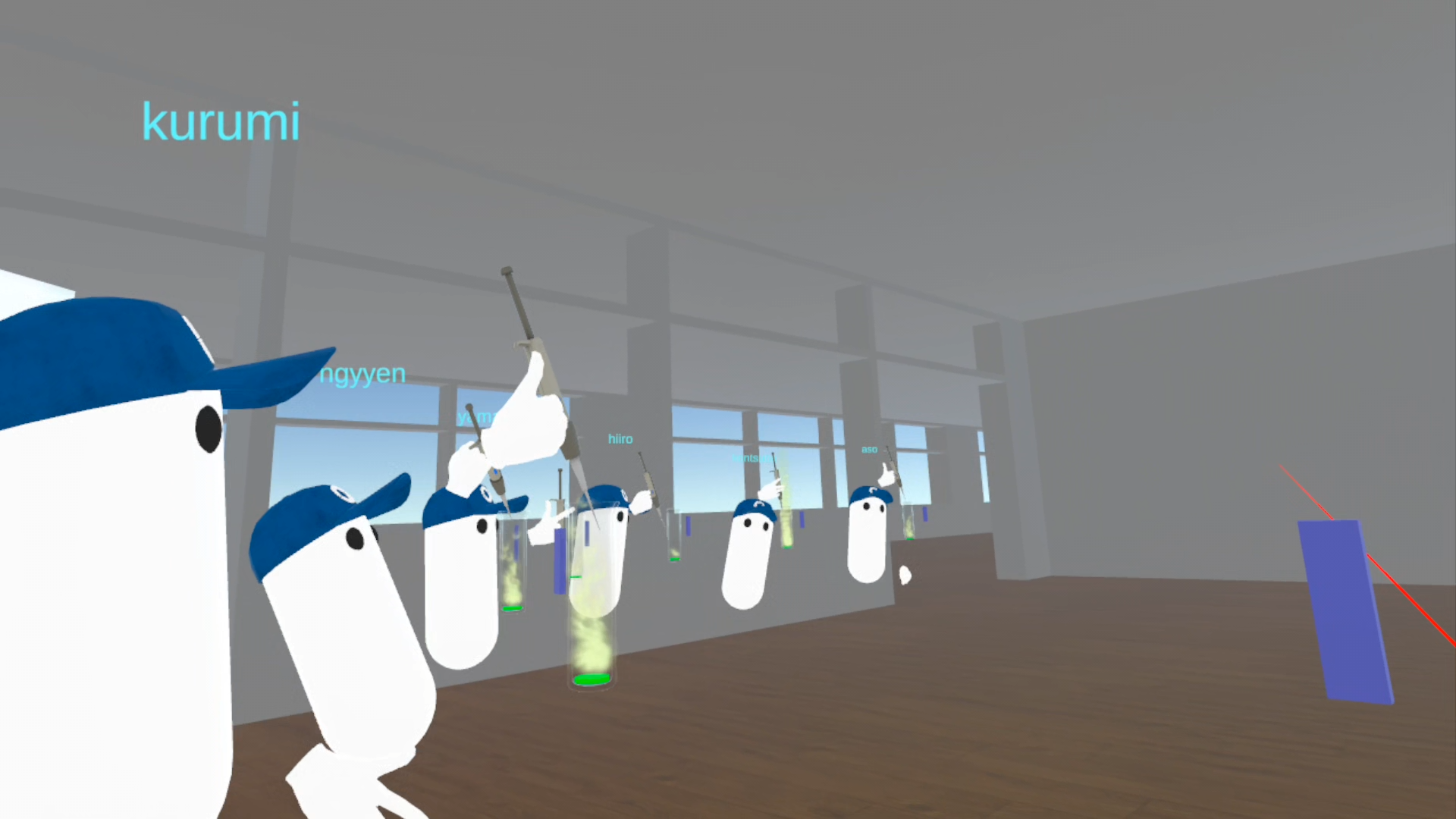}
  \end{center}
  \caption{System view from the teacher’s perspective.}
  \label{SystemsDemo}
  \vspace{-14pt}
\end{figure}

Fig.~\ref{SystemsDemo} shows an example of a lesson conducted in the proposed system from the teacher’s perspective. The teacher can interact with learning materials while communicating with students in real time, allowing for an immersive and interactive teaching experience.

\section{Evaluation}
\subsection{Experiment setting}
The experiment was conducted in a lab setting where students and teachers were in the same physical location and involved 10 undergraduate and graduate students from the same academic department. The participants were aged 21 to 24, with a gender ratio of 9 men to 1 woman. Additionally, 3 participants had previously used VR, while 7 had no experience. The experiment lasted one hour, including the briefing, and was conducted using the Meta Quest 3 VR headset. The experiment proceeded as follows: first, the participants launched the application and entered the VR classroom to familiarize themselves with the system. They then closed the application and took a break for 10 minutes. Afterward, they re-entered the classroom and conducted a science experiment. After the experiment, participants completed the System Usability Score (SUS)~\cite{SUS} and the Simulation Sickness Questionnaire (SSQ)~\cite{SSQ} questionnaires.


\subsection{Evaluation Results}
To evaluate the device load, we recorded the frame rate of the device used by the teacher during the test. The results showed that the average frame rate was 72.47(SD = 0.95 fps), indicating that the load was minimal.



We measured the end-to-end audio delay—the time from when Client A’s speech was transmitted to Client B, Client B heard it and responded immediately (by speaking), until that response reached Client A. The measurement results showed that the average audio response time between clients was 1.968 seconds (SD = 0.063 seconds). Given that the standard deviation was extremely small at 0.063 seconds, indicating high stability, this suggests that fluctuations in communication path delay (jitter) and variations in client processing load were minimal.

The SUS results of the proposed system are as follows. The average SUS score was 69.7, which corresponds to a grade of B, indicating good usability. However, scores ranged widely from 50.0 (D) to 92.5 (A), showing significant variation in user evaluations. In particular, while some users gave high ratings, others gave low ratings, suggesting that there are issues with User Interface (UI) consistency. The SSQ results of the proposed system are shown in Table~\ref{table:SSQ}. It can be seen that the sickness regarding nausea of the proposed system is minimal, while the one regarding oculomotor and disorientation is more significant. The total score is 12.8, indicating a minimal level of VR sickness~\cite{SSQ}.

\begin{table}[htbp]
\caption{Results of the Simulator Sickness Questionnaire.}
 \begin{center}
  \begin{tabular}{ccc} \hline
    Type & Average Score & SD\\ \hline
    Nausea & 3.18 & 4.77 \\
    Oculomotor & 17.69 & 17.78\\
    Disorientation & 10.83 & 15.21\\
    Total Score & 12.88 & 13.5\\ \hline
  \end{tabular}
 \end{center}
 \label{table:SSQ}
 \vspace{-12pt}
\end{table}
\section{Demo Movie for Review}
The demo video of our proposed system can be found at \url{https://youtu.be/1W4wGB9fCCI}.
\section{Conclusions}
In this study, a Virtual Reality system for conducting lessons in a virtual space in real time is developed and evaluated. Evaluation results showed that the proposed system provides stable communication with low latency and good usability, although improvements in UI consistency are needed. Furthermore, VR motion sickness was kept to a minimum. For future work, we plan to develop a wider variety of learning materials based on the proposed system, while also working to improve usability through enhancements to the UI and network system.

\bibliographystyle{IEEEtran}
\bibliography{reference}
\end{document}